\documentstyle[11pt]{article}

\def\ms{\widetilde{m}^2}
\def\H{\hat{H}}
\def\L{\hat{L}}
\def\Q{\hat{Q}}
\def\U{\hat{U}}
\def\D{\hat{D}}
\def\E{\hat{E}}
\def\O{\cal{O}}

\def\v{\vskip 5pt}

\title{
\begin{flushright}
\small
SINP/TNP/01-17\\
{\tt hep-ph/0108267}
\end{flushright}
  {\Large \bf  Supersymmetry as a physics beyond the \\ standard
    model \footnote{To be published in a Special Issue on High Energy
      Physics of the Indian Journal of Physics on the occasion of the
      75th anniversary of the journal. A shorter version with a
    different title will appear, as a plenary talk write-up, in the
    Proceedings of the XIVth DAE High Energy Physics Symposium,
    Hyderabad, India, 18-22 December, 2000.}}
}

\author{
{\sf Gautam Bhattacharyya}
\thanks{Electronic address: gb@theory.saha.ernet.in}
\\ [2.5mm]
{\small Saha Institute of Nuclear Physics, 1/AF Bidhan
Nagar, Kolkata 700064, India}
}

\date{}

\begin{document}

\maketitle

\begin{abstract}
  Here I briefly discuss why supersymmetry is considered a leading
  candidate of physics beyond the standard model.  I also highlight
  the salient features of different supersymmetry breaking models. A
  few other symmetries, broken or intact, asscociated with any
  realistic supersymmetric model are also identified. This write-up is
  too simple-minded for an expert on supersymmetry. It is basically
  intended for those who are busy in other areas of high energy
  physics.

\end{abstract}

\vskip 20pt

\setcounter{footnote}{0}
\renewcommand{\thefootnote}{\arabic{footnote}}
%%%%%%%%%%%%%%%%%%%%%%%%%%%%%%%%%%%%%%%%%%%%%%%%%%%%%%%%%%%%%%%%%%%%%

\section{Introduction}

Several experiments in the last few years have tested the standard
model (SM) of particle physics \cite{sm-books} to an unprecedented
accuracy. After the direct observations of the top quark and the
tau-neutrino, only the Higgs boson remains to be seen to bring the
search for SM particles to an end. Even though a 2.9 $\sigma$ signal
of a neutral Higgs boson weighing $\sim$ 115 GeV has of late been
announced by the LEP Collaborations at CERN \cite{lephiggs}, the
statistical significance nevertheless is too weak for the community to
accept it as a discovery of the Higgs.  As regards the parameters of
the SM, the total and partial widths of the $Z$ boson have been
measured at LEP to a {\em per mille} accuracy \cite{lepewwg}, and the
agreement with their SM predictions is unquestionable!  Although a few
discrepancies like the `$R_b$-$R_c$ crisis' and the `ALEPH four-jet
events' at LEP and the `high $Q^2$ anomaly' at the electron proton
collider HERA at DESY surfaced in the last few years, they all
disappeared with further accumulation of statistics and better
understanding of the detectors.  The inconsistency between the recent
muon $(g-2)$ measurement \cite{amuexp} and its SM prediction is not so
significant, particularly since the different SM computations
\cite{amusm} are not yet in complete agreement with one another.
Therefore we must acknowledge that the SM reigns supreme up to the
weak scale of order 100 GeV, successfully accounting for a huge set of
experimental data spanning over a wide range of energy.

\v

Then why do we at all bother about going beyond the SM? Actually a
number of theoretical prejudices suggest that the SM is a very good
effective description of Nature valid up to the weak scale. The SM
gauge group is arbitrary. We do not understand the electroweak
symmetry breaking mechanism, particularly, what makes the scalar
mass-square negative.  The SM cannot answer why there are only three
chiral generations. The SM has no reply to the question as to why the
weak scale is so light compared to the Planck scale. It also cannot
turn on gravity within its framework. Besides, the SM has as many as
18 free parameters. All these indicate that the SM may not be the
complete story.

\v

As regards the experimental data, even though it seems that everything
fits so well within the SM framework, there is some problem in the
paradise! At least in one place the SM is in genuine trouble -- this
is to explain the `non-zero' mass of a neutrino as suggested by the
neutrino oscillation data. Since the SM does not contain a
right-handed neutrino, it is not possible to account for the neutrino
mass within the renormalizable part of the SM Lagrangian.  Without
changing the SM field content, one can indeed generate a non-zero
neutrino mass through the following dimension-5 non-renormalizable
operator
\begin{equation}
{1\over \Lambda}\overline{L^C}LHH \leadsto
m_\nu \sim {{{\langle H\rangle}^2}
\over \Lambda},
\end{equation}
giving a Majorana mass suppressed by a high scale $\Lambda$.  Putting
$m_\nu \sim 0.1$ eV (the choice is motivated by the Super-Kamiokande
atmospheric neutrino data), we obtain $\Lambda \sim 10^{15}$ GeV,
which is tantalizingly close to the Grand Unification scale! This can
be interpreted \cite{giudice-talk} as a sort of indirect hint of a new
physical scale between $M_Z \sim 92$ GeV and $M_{\rm Pl} \sim 10^{19}$
GeV!

\v

All these suggest that at higher energies (i.e., shorter distances)
something beyond the SM becomes operative. But we cannot indulge
ourselves in wild thinking in new physics model building, since lots
of data are now there putting some kind of a discipline to our
imagination. According to majority `supersymmetry' \cite{susy-books,
  reviews} is the most attractive option to describe physics beyond
the SM.

\subsection{What is supersymmetry?}

Supersymmetry is a new space-time symmetry interchanging bosons and
fermions, i.e., it is a symmetry between states of different spin. As
an example, a spin-0 particle can be mapped to a spin-$1\over 2$
particle by a supersymmetry transformation. In supersymmetry, the
Poincare group is extended by adding two additional generators $Q$ and
$\bar{Q}$, which are anticommuting, to the existing $p$ (linear
momentum), $J$ (angular momentum) and $K$ (boost), such that
$\{Q,\bar{Q}\} \sim p$. Since the new symmetry generators are spinors,
not scalars, supersymmetry is not an internal symmetry. Years ago,
Dirac postulated a doubling of states by introducing antiparticle to
every particle to reconcile Special Relativity with Quantum Mechanics.
In Stern-Gerlach experiment, an atomic beam in an inhomogeneous
magnetic field was shown to split due to doubling of the number of
electron states into spin-up and -down modes. This indicated a
doubling of states with respect to angular momentum. So it is no
surprise \cite{hall} that $Q$ would cause a further splitting into
particle and superparticle ($f \stackrel{Q}{\rightarrow} f,
\widetilde{f}$). Since $Q$ is spinorial, the superpartners differ from
their SM partners in spin. The superpartners of fermions, called
sfermions, are scalars, and those of gauge bosons, called gauginos,
are fermions. They are clubbed together to form supermultiplets. The
two irreducible supermultiplets which are used to construct the
supersymmetric standard model are the `chiral' and the `vector'
supermultiplets. The chiral supermultiplet contains a scalar $\phi$
and a 2-component Majorana fermion $\psi$. The vector supermultiplet
contains a gauge field $A_\mu$ and a 2-component Majorana fermion
$\lambda$ (gaugino).  The generic features of a supermultiplet are:
\begin{itemize}
\item There is an equal number of bosonic and fermionic degrees of
  freedom in a supermultiplet.
\item Since $p^2$ commutes with $Q$, the bosons and fermions in a
  supermultiplet are mass degenerate.
\end{itemize}

\v

The minimal supersymmetric standard model (MSSM) is the supersymmetric
extension of the SM which has the minimal particle content: two
complex Higgs doublets (the SM has only one) and their superpartners,
the SM fermions and gauge bosons and their superpartners. The quartic
scalar coupling is related to the gauge coupling; in that sense it has
one less parameter than in the SM when supersymmetry is unbroken.
There is an additional parameter $\mu$ which controls the mixing
between the Higgs supermultiplets.  It should be noted that
supersymmetry has to be broken in order to explain the non-observation
of any superparticle to date, and this introduces additional
parameters. The scalar and gaugino masses, the trilinear scalar
couplings ($A$ parameters) and the bilinear Higgs mixing parameter
($B_\mu$) are these additional parameters. In the absence of a precise
knowledge about supersymmetry breaking, these parameters are in
general free and unrelated to one another. The number of such
parameters will be counted later. The MSSM field contents are
summarised in Table 1.

\begin{table}[h]
\small
\caption{{\small \sf The particles in the minimal supersymmetric standard
model. Overhead `tilde' indicates superpartner.}}
\begin{center}
\begin{tabular}{cccc}
\hline
\hline
Particles/superparticles
 & spin 0 & spin 1/2 & $SU(3)\otimes SU(2) \otimes U(1)$ \\
\hline
\hline
 & & & \\
leptons, sleptons ($L$) & $(\widetilde{\nu}, \widetilde{e}_L)$ &
 $({\nu}, {e}_L)$ & ($1, 2, -1/2$) \\
(in 3 families) ($E^c$) & ${\widetilde{e}_R}^*$ & $e^c_L$ & (1, 1, 1)
\\
 & & & \\
\hline
 & & & \\
quarks, squarks ($Q$) & $(\widetilde{u}_L, \widetilde{d}_L)$ &
 $({u}_L, {d}_L)$ & (3, 2, 1/6) \\
(in 3 families) ($U^c$) & ${\widetilde{u}_R}^*$ & $u^c_L$ &
($\bar{3}, 1, -2/3$) \\
 ($D^c$) & ${\widetilde{d}_R}^*$ & $d^c_L$ &
($\bar{3}, 1, 1/3$) \\
 & & & \\
\hline
  & & & \\
Higgs, higgsinos ($H_u$) & $(H_u^+, H_u^0)$ &
 $(\widetilde{H}_u^+, \widetilde{H}_u^0)$ & (1, 2, 1/2) \\
 ($H_d$) & $(H_d^0, H_d^-)$ &
 $(\widetilde{H}_d^0, \widetilde{H}_d^-)$ & ($1, 2, -1/2$) \\
 & & & \\
\hline
\hline
Particles/superparticles
& spin 1 & spin 1/2 & $SU(3) \otimes SU(2) \otimes U(1)$ \\
\hline
\hline
 & & & \\
gluon, gluino & $g$ & $\widetilde{g}$ & (8, 1, 0) \\
$W$ bosons, winos & $W^\pm, W^0$ & $\widetilde{W}^\pm, \widetilde{W}^0$ &
(1, 3, 0) \\
$B$ boson, bino & $B^0$ & $\widetilde{B}^0$ &
(1, 1, 0) \\
 & & & \\
\hline
\hline
\end{tabular}
\end{center}
\end{table}

\subsection{Why supersymmetry?}

\begin{itemize}
\item {\sf Supersymmetry protects the electroweak hierarchy from
destabilizing divergences.}

The infamous `hierarchy problem', i.e., why $M_{\rm Pl} \gg M_W$, or
equivalently, $G_N \ll G_F$, is the main motivation behind the
introduction of supersymmetry \cite{witten,dimo-georgi,quad_div}. It
should be noted that the fermion masses are protected by chiral
symmetry and the gauge boson masses by gauge symmetry. But there is no
such protection mechanism for a scalar. In the SM, $m_H \sim M_W$,
both being proportional to the Higgs vacuum expectation value (VEV).
The quantum correction to the Higgs mass is $\delta m^2_H =
{\O}(\alpha/\pi)\Lambda^2$, where $\Lambda$ is the ultraviolet cut-off
scale reflecting the appearance of new physics. It is nevertheless
possible to bring down the Higgs mass to weak scale by adjusting the
counter-term to cancel this large quantum correction. But this
adjustment has to be done order by order and such fine-tuning of one
part in $10^{17}$ to keep the Higgs mass at the weak scale makes the
theory sick. The requirement of such an unnatural cancellation is at
the root of the hierarchy problem. We note at this point that quantum
corrections to the Higgs mass from a bosonic loop and a fermionic loop
have {\em opposite} sign.  So if the couplings are same and the boson
is mass degenerate with the fermion, the correction would vanish! What
can be a better candidate than supersymmetry to do this job? For every
particle supersymmetry provides a mass degenerate partner differing by
spin $1\over 2$.  However, the cancellation is not exact because in
real world supersymmetry is broken. But it has the virtue that it
makes the Higgs mass quantum correction milder: $\delta m^2_H =
{\O}(\alpha/\pi)\delta m^2$, where $\delta m^2$ is the splitting
between partners and spartners. Clearly, $\delta m^2_H < m_H^2$, if
$\delta m^2 < 1~{\rm TeV}^2$.

\item {\sf Supersymmetry leads to unification of gauge couplings.}

  This is another motivation \cite{uni}. In the SM, when the gauge
  couplings are extrapolated to high scale, with their low energy
  measurements as input values for the calculation of their running,
  they do not meet at a single point. In supersymmetry, they do, at a
  scale $M_{\rm GUT} \sim 2 \times 10^{16}$ GeV. The requirement is
  that the superparticles weigh around 1 TeV. In other words,
  supersymmetry has the right particle content to ensure that the
  gauge couplings can unify. This could be just an accident, but may
  be taken as a strong hint in favour of the Grand Unified theories
  (GUT). In the same spirit, we may expect that other supersymmetric
  couplings and soft masses also unify.

\item {\sf Supersymmetry prefers a heavy top quark, and the top is
indeed heavy.}

Supersymmetry has two Higgs doublets. Due to heavy top quark induced
radiative correction, the mass-square of one of the Higgs bosons, the
one that couples to the up quark, starting from a positive value in
the ultraviolet becomes negative in the infrared triggering
electroweak symmetry breaking (EWSB). In the SM, the negative value of
scalar mass-square is completely {\em ad hoc} and is put in by hand to
ensure EWSB. In supersymmetry it is the heavy top quark that induces
the flip of sign of one scalar mass-square. Also, it is quite amazing
that it is the Higgs that becomes tachyonic during renormalization
group running.

\item {\sf Supersymmetry prefers a light Higgs, and the Higgs is expected
  to be light.}

Supersymmetry predicts that the lightest neutral Higgs is lighter than
$M_Z$ at tree level. Due to the heavy top quark induced radiative
correction it can at most go up to $\sim$ 150 GeV. Even though the
Higgs has not yet been found, the precision electroweak data suggest
that it should lie within 195 GeV at 95\% CL, while the direct lower
limit from LEPII is $m_H > 113.5$ GeV at 95\% CL.

Also, if $m_H \sim 115$ GeV (the hinted value!), the SM effective
potential becomes unstable above $\Lambda \sim 10^6$ GeV. This can be
rectified only by invoking a `supersymmetry-seeming' set of new
physics \cite{ellis_ross}.

\item {\sf Supersymmetry is a decoupling theory.}

  Supersymmetry decouples from precision measurements of the $Z$-pole
  observables. In other words, supersymmetry contributions vanish as
  superparticle masses are pushed to infinity. On the contrary, simple
  technicolour models do not decouple and they are ruled out from
  precision electroweak measurements.

\item {\sf Supersymmetry provides a cold dark matter candidate.}

  If the gluino, the superpartner of gluon, is around 1 TeV, then the
  GUT boundary conditions imply that the lightest supersymmetric
  particle (LSP) is around 100 GeV. Consistency with cosmology demands
  that the LSP is colour and electrically neutral \cite{lsp_cosm}.
  The LSP interacts weakly with other particles and has properties
  which make it a natural candidate for cold dark matter of the
  universe. However, one necessary condition for this is that the LSP
  is stable. We will see later that in some supersymmetric models the
  LSP decays into the SM particles.

\item {\sf Supersymmetry provides a framework to turn on gravity.}

  Using supersymmetry as a local symmetry leads to supergravity
  models. This is how gravity can be unified with all other
  interactions.

\item{\sf Supersymmetry can explain the neutrino oscillation data.}

Supersymmetric models with broken lepton numbers can reproduce the
neutrino masses and mixing angles compatible with the recent neutrino
oscillation data.

\end{itemize}

\subsection{Where are the superparticles?}

Not a single superparticle has been found to date! The lower limits
from direct searches at colliders on the masses of generic sleptons
($\widetilde{l}$) and squarks
($\widetilde{q}$), gluino ($\widetilde{g}$), lightest chargino
($\widetilde{\chi}_1^\pm$) and the lightest neutralino
($\widetilde{\chi}_1^0$) are
\begin{eqnarray}
\label{bound}
m_{\widetilde{l}} & > &\sim 100~{\rm GeV} ~~({\rm LEPII}) , \nonumber \\
m_{\widetilde{q}} \sim m_{\widetilde{g}} & > & \sim 300~{\rm GeV}
~~({\rm Fermilab ~ Tevatron}), \nonumber \\
m_{\widetilde{\chi}_1^\pm} & > & \sim 100~{\rm GeV} ~~({\rm LEPII}), \\
m_{\widetilde{\chi}_1^0} & > & \sim 30~{\rm GeV} ~~({\rm using ~GUT
~relations}). \nonumber
\end{eqnarray}
Eq.~(\ref{bound}) shows that supersymmetry is not only broken, it is
very badly broken!

\v

Since supersymmetry has lots of unknown parameters, translating
experimental data into allowed/excluded multidimensional supersymmetry
parameter space is a complicated job. First we have to decide what we
are expecting to observe, and then we have to design our devices to
detect them. Superparticles are expected to be produced in pairs, and
if the LSP is stable, then the missing energy the latter carries
constitutes the characteristic signature for supersymmetry search. For
detailed and instructive discussions on the production and decay of
superparticles in different colliders, see, for example, the reviews
by Martin (hep-ph/9709356), Gunion (hep-ph/9704349), Tata
(hep-ph/9706307) and Dawson (hep-ph/9612229).  For a review on how to
look for supersymmetry in the next linear collider (NLC), see the
recent Snowmass report \cite{snowmass}.

\v

For indirect constraints on supersymmetry parameter space from
electroweak precision data, $b \to s \gamma$, $(g-2)_\mu$, cosmology,
etc., the readers are referred to, for example, \cite{altarelli},
\cite{ellis}, \cite{boer}. One recent analysis \cite{altarelli}
suggests that the quality of electroweak fit improves with light
superpartners!  Admitting sneutrinos in the range 55-80 GeV and
charged sleptons marginally above their experimental lower limit fit
the data better than just the SM alone!

\section{Supersymmetry breaking}

Let us first consider the spontaneous breaking of supersymmetry.
There is a difference between an internal symmetry breaking and
supersymmetry breaking. The latter requires $\langle 0|H|0\rangle >
0$, where $H = {1\over 4}\sum Q_\alpha^2$. Consider a situation in which
$V(\phi) = 0$ for $\phi = v \neq 0$, and $V(0) > 0$, where $V$ is the
scalar potential, and $\phi$ is a scalar field which obtains a VEV
$v$ (e.g., consider the so called Mexican Hat
potential of the SM). Here gauge symmetry is broken but supersymmetry
is unbroken. Now consider a different situation in which $V$ is
symmetric about $\phi = 0$, with $V (0) = V_{\rm min} = E > 0$. Here
gauge symmetry is unbroken, but supersymmetry is broken.

\v

There two important points \cite{witten}:
\begin{itemize}
\item When supersymmetry is spontaneously broken, a massless Goldstone
fermion is generated. Remember, it is not simply a massless fermion
(like neutrino), Goldstone fermion is a massless fermion that is
created from vacuum by the supersymmetry current:
\begin{equation}
\langle 0|S_{\mu\alpha}|\psi_\beta\rangle = (\gamma_\mu)_{\alpha\beta}
f; ~~ Q_\alpha = \int d^3x ~S_{0\alpha}; ~~E = f^2,
\end{equation}
where $f$ is the coupling of the supercurrent to the Goldstone fermion
and $E$, as stated previously, is the vacuum energy density.

\item If $f = 0$ at tree level, it remains zero at all order.
\end{itemize}

But if supersymmetry is spontaneously broken at tree level in the
observable world, it immediately leads to a disaster!  The reason is
the presence of a mass sum rule \cite{ferrara}
\begin{equation}
{\rm STr} M^2 = \sum_{J=0}^1 (-)^{2J} (2J + 1) {\rm Tr} M_J^2 = X,
\end{equation}
where $J$ is the spin of the particle, and $X$ corresponds to the
trace of the group generators. $X = 0$ for any U(1) trace-anomaly free
models, e.g., the supersymmetric standard model. This sum rule creates
a major threat, since it predicts the existence of a charge $2/3$
squark not heavier than the lightest charge $2/3$ quark, and/or, a charge
$-1/3$ squark not heavier than the lightest charge $-1/3$ quark
\cite{dimo-georgi}. In view of the non-observation of any
superparticle to date, both results are experimentally ruled out.  One
way to tackle this problem is to admit explicit breaking of
supersymmetry. At the same time we must ensure that the hierarchy
between the Planck scale and the weak scale is not destabilized by
such an action, because that was after all the motivation behind
introducing supersymmetry in the first place.  The terms which break
supersymmetry explicitly but do not regenerate the quadratic
divergences are called `soft terms'. In fact, supersymmetry breaking
is implemented by admitting explicit mass terms for the scalars in the
chiral multiplets and gauginos in the vector multiplets, and
additionally by introducing the bilinear Higgs mixing ($B_\mu$ term)
and triliear scalar interaction ($A$ terms) in the soft Lagrangian.
The mass dimension of the soft terms in the Lagrangian must be $\leq
3$. We must however keep in mind that these soft terms are not quite
arbitrary, as otherwise it would be difficult to satisfy many
experimental constraints, e.g., the suppression of flavour-changing
neutral currents.

\v

But what is the origin of these soft terms? The usual prescription is
the following: Supersymmetry is spontaneously broken in a `hidden
sector' which has no (or very small) interactions with the `visible
sector' supermultiplets. But the two sectors share some common
interactions which mediate the information of supersymmetry breaking
from the hidden sector to the observable world.  The result is the
appearance of calculable soft terms in the observable sector
Lagrangian. This prescription leads to different mediation mechanisms
of supersymmetry breaking.  These mechanisms differ in the way the
soft masses are generated and related to one another.

\subsection{Supergravity}

In supergravity (SUGRA) models \cite{sugra}, global supersymmetry is
promoted to local supersymmetry.  Supersymmetry is broken in the
hidden sector and the message is transmitted to the observable sector
by $(1/M_{\rm Pl})$-suppressed operators.  Analogous to the Higgs
mechanism in the SM, there is a super-Higgs mechanism operative here
in which the gravitino `eats up' the Goldstino and becomes massive.
The gravitino mass ($m_{3/2}$) is therefore a `hard' parameter which
has its origin in the hidden sector. The essential points are:

\begin{itemize}

\item Denoting the supersymmetry breaking scale by $\sqrt{F}$, one can
  express $m_{3/2} \sim F/M_{\rm Pl}$. For $m_{3/2} \sim 1$ TeV, we
  obtain $\sqrt{F} \sim 10^{11}$ GeV.

\item The soft scalar masses $\ms \propto G_N \propto 1/M_{\rm
Pl}^2$. Therefore, $\ms \sim F^2/M_{\rm Pl}^2$, i.e. $\widetilde{m} \sim
m_{3/2} \sim 1$ TeV.

\item The $\mu$ problem is solved by the Giudice-Masiero mechanism
  \cite{giudice-masiero}, in which the $\mu$-term is generated only at the
  time of supersymmetry breaking as a consequence of the observable
  sector's interaction with the hidden sector. The Higgs mixing part is
\begin{equation}
\int d^2\bar{\theta} z^* \H_d \H_u/M_{\rm Pl},
\end{equation}
where $z$ is a hidden sector spurion field which obtains an $F$-term
vev $F_z {\bar{\theta}}^2$ ($\bar{\theta} =$ Grassmann variable).
$F_z$ breaks supersymmetry, and at the same time generates $\mu \sim
F_z/M_{\rm Pl} \sim 1$ TeV.

\item There are 4 free parameters and 1 sign: the common scalar and
  gaugino masses $m_0$ and $M_{1/2}$ respectively, the common
  trilinear $A$ parameter, the bilinear $B_\mu$ parameter (all at the
  unification scale), and the sign of the $\mu$ parameter.

\item The lightest supersymmetric particle (LSP) is the lightest
neutralino. The supersymmetry search strategies are based on the
characteristic missing energy signature.

\end{itemize}

A closer look to the soft-breaking terms reveals that the Kahler
potential generates a mass term of the form $C_{ab}
{\phi_a}^\dagger{\phi_b}$, where $C_{ab} = h_{ab} F^2/M^2_{\rm Pl}$,
with $a$ and $b$ being the generation indices. There is no guarantee
that the coupling $h$ will be flavour digonal. The real source of this
Planck scale suppressed coupling lies in integrating out some of the
Planck scale states which may couple to both hidden and observable
sectors. Unless these couplings are flavour diagonal at high energy,
we cannot control the magnitudes of the flavour off-diagonal couplings
at low energy. On the other hand, experimental constraints (like
$\Delta m_K$, $\mu \to e\gamma$, etc.)  at low energy imply that
flavour changing neutral currents are highly suppressed.  The lack of
explanation as to why these couplings would be flavour diagonal (i.e.
$C_{ab} = C \delta_{ab}$) at high scale within the framework of
supergravity gives rise to the `supersymmetric flavour problem'.

\subsection{Gauge mediation}

Gauge mediated supersymmetry breaking (GMSB) models \cite{gmsb} have
been formulated primarily for the purpose of removing the
supersymmetric flavour problem. Here gauge interactions rather than
gravity are used to transmit the message of supersymmetry breaking. In
this framework, there is a hidden sector where supersymmetry is
broken, and in between the hidden and observable sectors there is a
messenger sector. The messenger particles have couplings with standard
gauge bosons and are aware of supersymmetry breaking since they have
direct interactions with hidden sector fields as well. Consider the
superpotential: $W = \lambda X M \overline{M}$, where $X$ is a hidden
sector field and $M$ and $\bar{M}$ are messenger sector fields which
could be a 5 and $\bar{5}$ or a 10 and $\overline{10}$ of SU(5). $\langle
X\rangle$ breaks $U(1)_R$ symmetry (discussed later) and $\langle F_X
\rangle \equiv F$ breaks supersymmetry. The messenger scalars are then
split as $m^2_\pm = M^2 \pm F$. The main features of GMSB models are:
\begin{itemize}
\item Gaugino mass is generated at one-loop at the messenger scale
$M$. One obtains
\begin{equation}
m^i_{\lambda} \sim {\alpha_i \over {4\pi}} {{F
\langle X \rangle} \over{M^2}},
\end{equation}
i.e., non-zero gaugino mass requires the breaking of both $R$ symmetry
and supersymmetry.

\item Scalar masses are generated at two-loop order at the scale $M$.
\begin{equation}
\ms = 2n (F/M)^2 \sum_{i=1}^3 C_i
\left({{\alpha_i^2(M)}\over{4\pi}}\right),
\end{equation}
where $n$ is the number of messenger multiplets, and $C_i$'s are the
quadratic Casimir coefficients for the different representations to
which the scalars belong. Clearly, squark masses being proportional to
strong coupling constant are heavier than sleptons, since for the
latter $i$ cannot be 3. Also note that even though squark masses are
generated in two-loop, and gaugino masses in one-loop, the former
appear as squared masses while the latter linearly, and hence both are
of the same order.

\item The $\mu$ term is generated at one-loop but $B_\mu$ in scalar
  potential is generated at two-loop. Still, this is very difficult to
  achieve. The `$\mu$ problem' is rather serious in GMSB models
  \cite{gmsb-mu}.  The trilinear $A$ parameter is generated at
  two-loop, and at electroweak scale is not very large.

\item The messenger scale $M$ can be as low as 100 TeV. This means
that a squark mass of $\sim$ 100 GeV is consistent with a $\sqrt{F}
\sim 100$ TeV. Recall that in SUGRA scenario $\sqrt{F}$ is several
orders of magnitude larger. For this reason the GMSB models are often
cited as the `low scale supersymmetry breaking models'.

\item The gravitino is superlight: $m_{3/2} \sim F/M_{\rm Pl} \sim
0.1$ eV or so! Gravitino emission from hard photons or from selectrons
constitutes the `smoking gun' signals.

\end{itemize}

\subsection{Anomaly mediation}

The motivation is again to solve the flavour problem.  In this
scenario \cite{amsb} the hidden sector is in one brane and the
observable sector is in a different brane. Supersymmetry breaking
takes place in the hidden sector and is transmitted to the observable
sector via superconformal anomaly. Unlike in the SUGRA scenario, here
there is no Kahler coupling between the hidden and the observable
sectors, and this way the soft terms become ultraviolet insensitive.
The `flavour problem' is thus taken care of.  Superconformal anomaly
gives rise to a non-zero trace of energy-momentum tensor, and as a
result soft masses are generated being proportional to the beta
function of the corresponding interaction.  The main features of this
anomaly mediated supersymmetry breaking (AMSB) models are:

\begin{itemize}
\item  Gaugino masses are proportional to the gauge beta functions:
$\widetilde {M}_i \propto \beta_i m_{3/2}$. More precisely,
\begin{eqnarray}
\widetilde{M}_3 =  -{{3\alpha_s}\over{4\pi}} m_{3/2};~
\widetilde{M}_2 = {{\alpha}\over{4\pi\sin^2\theta_w}} m_{3/2};~
\widetilde{M}_1 = {{11\alpha}\over{4\pi\cos^2\theta_w}} m_{3/2}.
%\label{}
\end{eqnarray}
This means $\widetilde{M_3}:\widetilde{M_2}:\widetilde{M_1} =
1:-0.1:-0.3$, i.e., the Wino is the lightest neutralino. This should
be compared with the SUGRA gaugino mass relation:
$\widetilde{M_3}:\widetilde{M_2}:\widetilde{M_1} =
1:0.3:0.17$.

\item The soft scalar masses are given by
\begin{eqnarray}
\ms = -0.25 \left(\frac{\partial{\gamma}}{\partial{g}} \beta_g
+ \frac{\partial{\gamma}}{\partial{y}} \beta_y \right)m^2_{3/2},
%\label{}
\end{eqnarray}
where $\gamma = c_0 g^2 + d_0 y^2$, $\beta_g = -b_0 g^3$, $\beta_y =
y(e_0 y^2 + f_0 g^2)$, with $g$ and $y$ being the gauge and Yukawa
couplings. Since the SU(3) gauge beta function is negative while the
SU(2) and U(1) gauge beta functions are positive, it immediately
follows from the above mass relation that the sleptons are
`tachyonic'. Phenomenologically, one tackles this problem by putting
in an universal $m_0^2$ by hand to all scalar masses. A convincing
theoretical reasoning regarding the origin of $m_0^2$ is still
lacking.

\item AMSB soft masses are renormalisation group invariant. Hence they
are completely determined by their known low energy gauge and Yukawa
couplings and hence make no reference to their high energy boundary
conditions. However, the universal $m_0$, which is thrown in to solve
the tachyonic slepton problem, spoils this invariance.

\item The spectrum is defined in terms of 3 parameters and 1
sign: $m_{3/2}$, $m_0$, $\tan \beta$, and the sign of $\mu$.

\end{itemize}
The near degeneracy between the lighter chargino and the wino
dominated neutralino LSP is the issue that one employs to constitute
the clinching test of this scenario. The chargino will decay into LSP
and a `very soft' pion and this decay will give rise to a displaced
vertex. Triggering such events though is not an experimentally easy
task!

\section{The parameters in a general supersymmetric model}
In a general supersymmetric model, where we do not assume any
particular mediation mechanism and do not impose any GUT conditions,
the soft parameters are not related to one another. In this section,
we will see how a general supersymmetric model can be parametrized
\cite{feyn-rules}. We will also count the number of parameters
required for this purpose.

\v

First consider the superpotential, written in terms of the `chiral
superfields', as
\begin{eqnarray}
W = \sum_{ij} \left(h_e^{ij} \L_i \H_d \E_j^c + h_d^{ij} \Q_i \H_d \D_j^c +
h_u^{ij} \Q_i \H_u \U_j^c\right) +
\mu \H_d \H_u.
\label{w}
\end{eqnarray}
Above, the sum is over the different generations.  $\H_d$ and $\H_u$
are the two Higgs doublet superfields. The former gives masses to
down-type quarks and charged leptons and the latter gives masses to
up-type quarks. $\L$ and $\Q$ are lepton and quark doublet
superfields; $\E^c$, $\D^c$ and $\U^c$ are the singlet charged lepton,
down quark and up quark superfields respectively. $h_e$, $h_d$ and
$h_u$ are the Yukawa couplings and $\mu$ is the Higgs mixing
parameter. The usual convention is to put a `hat' over a superfield,
and without that `hat' the symbol represents the scalar component
within that superfield.

\v

The Lagrangian is given by
\begin{eqnarray}
-L = \sum_i \left|\frac{\partial {W}}{\partial {\phi_i}}\right|^2
+ \sum_{ij}\frac{\partial^2{W}}{\partial \phi_i \partial \phi_j}
\psi_i \psi_j
+
\frac{1}{2} \sum_\alpha |D_\alpha|^2 + \sum_{ij\alpha} \sqrt{2}
g_\alpha \psi_i (T^\alpha)^i_j \phi_j^* \lambda_\alpha,
\label{l}
\end{eqnarray}
where $\phi_i$ and $\psi_i$ the generic scalar and fermion fields
within the $i$th chiral multiplet, and $\lambda_\alpha$ represents the
gaugino which is a Majorana fermion in the vector multiplet with
$\alpha$ as the gauge group index.  The $D$ term is given by $D_\alpha
= -g_\alpha \phi_i (T^\alpha)^i_j \phi_j^*$.

\v

The soft breaking terms are given by ($i,j$: generation indices,
$\alpha$: gauge group label)
\begin{eqnarray}
-L_{\rm soft} & = &
\sum_{ij} \widetilde{m}_{ij}^2 \phi_i^*\phi_j +
\sum_{ij} \left(A_e^{ij} L_i H_d E_j^* + A_d^{ij} Q_i H_d D_j^* +
A_u^{ij} Q_i H_u U_j^*\right) \nonumber \\
& + & m_{H_d}^2 |H_d|^2 + m_{H_u}^2 |H_u|^2 +
(B_\mu H_d H_u + {\rm h.c.}) +
{1\over 2}\left(\sum_\alpha \widetilde{M}_\alpha \lambda_\alpha
\lambda_\alpha + {\rm h.c.}\right).
\label{soft}
\end{eqnarray}

\subsection{Counting parameters}
Let us now count the total number of real and imaginary parameters in
a general supersymmetric model with 2 Higgs doublets
\cite{dimo_sutter}.  Each Yukawa matrix $h_f$ in Eq.~(\ref{w}) has 9
real and 9 imaginary parameters, and there are 3 such matrices.
Similarly, each $A_f$ matrix in Eq.~(\ref{soft}) has 9 real and 9
imaginary parameters, and again there are 3 such matrices.  The scalar
mass square $\widetilde{m}_{ij}^2$ can be written for 5
representations: $Q, L, U^c, D^c, E^c$. For each represenation, the
$(3 \times 3)$ hermitian mass square matrix has 6 real and 3 imaginary
parameters. Finally, we have 3 gauge couplings (3 real), 3 gaugino
masses (3 real and 3 imaginary), $\mu$ and $B_\mu$ parameters (2 real
and 2 imaginary), $(m_{H_u}^2, m_{H_d}^2)$ (2 real), and $\theta_{\rm
  QCD}$ (1 real).  Summing up, there are 95 real and 74 imaginary
parameters. Are all of them physical? The answer is: No! If we switch
off the Yukawa couplings and the soft parameters, i.e., turn on {\em
  only} gauge interactions, there is a global symmetry, given by
\begin{equation}
 G_{\rm global} = U(3)^5 \otimes U(1)_{\rm PQ} \otimes U(1)_R.
\label{G}
\end{equation}
The Peccei-Quinn (PQ) and $R$ symmetries will be discussed in the next
two subsections. $U(3)^5$ symmetry means that a unitary rotation to the 3
generations for each of the 5 representations leaves the physics
invariant. However, this unitary symmetry is broken. Once a symmetry
is broken, the number of parameters required to describe the symmetry
transformation can be removed. For example, when a U(1) symmetry is
broken, we can remove one phase. Since a U(3) matrix has 3 real and 6
imaginary parameters, we can remove 15 real and 30 imaginary
parameters from the Yukawa matrices once $U(3)^5$ is broken. As we
will see, the PQ and $R$ symmetries are also
broken. So we can remove 2 more imaginary parameters. But even when
all the Yukawa couplings and soft parameters are turned on, there is a
still a global symmetry
\begin{equation}
G'_{\rm global} = U(1)_B \otimes U(1)_L,
\end{equation}
where $B$ and $L$ are baryon and lepton numbers. Hence we can remove
{\underline {not}} 32 but {\underline {only}} 30 imaginary parameters.
Thus we are left with 95 $-$ 15 = 80 real and 74 $-$ 30 = 44
imaginary, i.e., a total of 124 independent parameters. The SM had
only 18. So supersymmetry gifts us \underline{106} more!  In the SM we
had only one CP violating phase. Now we have 43 \underline{new} phases
which are CP violating!

\subsection{Peccei-Quinn (PQ) symmetry}
If we put $\mu = 0$ in Eq.~(\ref{w}), $W$ is invariant under the
following global U(1) transformation:
\begin{eqnarray}
\H_{d(u)} \to e^{i\alpha} \H_{d(u)}, \Q(\L) \to e^{-i\alpha} \Q(\L),
\E^c \to \E^c, \D^c \to \D^c, \U^c \to \U^c.
\label{pq}
\end{eqnarray}
This is a PQ symmetry, and it is preserved as long as $\mu = 0$. Note
that if $\mu = 0$, the $B_\mu$ parameter in Eq.~(\ref{soft}) is also
zero.  Now consider the scalar minimization condition: $B_\mu = m_A^2
\sin 2\beta$, where $\tan\beta = v_u/v_d = \langle H_u\rangle/\langle
H_d\rangle$. If $B_\mu = 0$, two cases may arise: (1) either $v_d = 0$
or $v_u = 0$, i.e., some quarks/charged leptons are massless, (2) $m_A
= 0$. Both cases are experimentally ruled out. Hence $\mu \neq 0$, and
the PQ symmetry is broken. Consistency with phenomenology requires
$\mu$ to be within 1 TeV.

\v

But $\mu$ is a superpotential parameter and hence there is no reason
for it to be zero in the limit of exact supersymmetry. In fact it
could in principle be as high as the GUT or Planck scale. Then the
question is what makes it to weigh in the ball-park of other
supersymmetry breaking masses? This is the origin of the so called
`$\mu$ problem'.

\subsection{$R$ symmetry}

Under this global U(1) symmetry, the superpotential $W \to W'
=e^{2i\alpha} W$. This can be arranged by, for example, by the following
choice:
\begin{eqnarray}
\H_{d(u)} \to e^{i\alpha} \H_{d(u)}, \Q(\L) \to e^{i\alpha} \Q(\L),
\E^c \to \E^c, \D^c \to \D^c, \U^c \to \U^c.
\label{r}
\end{eqnarray}
Note, $L = \int d^2\theta W$ is invariant under this rotation, where
$\theta$ is the Grassmann variable. Therefore, $d\theta \to
e^{-i\alpha} d\theta$ and $\theta \to e^{i\alpha} \theta$ (to ensure
$\int d\theta \theta = 1$). Since a chiral superfield ($\hat\phi$) can
be written in terms of its scalar ($\phi$) and fermion ($\psi$)
components as $\hat{\phi} = \phi + \psi \theta + F \theta^2$ (where
$F$ is the auxiliary parameter), it is clear that $\phi$ carries the
same $R$ charge as $\hat{\phi}$ and $\psi$ carries one unit less. It
immediately follows that for the Lagrangian in Eq.~(\ref{l}) to remain
invariant, the gaugino picks up a non-zero $R$ charge: $\lambda_\alpha
\to e^{i\alpha}\lambda_\alpha$. It then follows from Eq.~(\ref{soft})
that the gauginos are massless when $R$ symmetry is exact. Also, the
trilinear $A$ terms are vanishing in the same limit. So $R$ symmetry
has to be broken for the construction of a realistic supersymmetric
model.

\subsection{CP violation in supersymmetry}
For simplicity, let us consider the MSSM with common scalar and
gaugino masses ($\widetilde m$ and $\widetilde{M}$ respectively). We
also assume $A_f = A h_f$. In this framework, there are 4 additional
(compared to the SM) CP violating phases \cite{dugan,dimo,nir}. These
are contained in the 4 parameters $\widetilde{M}$, $A$, $B_\mu$ and
$\mu$. But we will see that only 2 of the 4 phases are physical. To
realise this, let us treat the above 4 parameters (appearing in
Eqs.~(\ref{w}) and (\ref{soft})) as spurions whose VEVs break the PQ
and $R$ symmetries.  They are thus assigned the following PQ and $R$
charges to compensate those of the fields described in Eqs.~(\ref{pq})
and (\ref{r}):
\begin{equation}
\mu : (-2, 0) ~~~~B_\mu: (-2, -2) ~~~~ A: (0, -2) ~~~~
\widetilde{M}: (0, -2),
\end{equation}
where, for each parameter, the first number denotes the PQ charge and
the second number the $R$ charge. The arguments of only those
combinations will give independent phases which have no net PQ and $R$
charges. There are 2 such independent phases:
\begin{equation}
\phi_A = {\rm Arg}~(A^*\widetilde{M}), ~ \phi_B = {\rm
  Arg}~(\widetilde{M}\mu B_\mu^*).
\end{equation}

\v

These phases contribute to the electric dipole moment of the neutron
as
\begin{equation}
d_N \sim 2 \left({100~{\rm GeV}}\over \widetilde{m}\right)^2 \sin
\phi_{A,B} \times 10^{-23} e~{\rm cm} ~\cite{paban},
\end{equation}
which should be compared with $d_N^{\rm exp} < 6.3 \times 10^{-26} e$
cm \cite{harris}. Generically we may expect $\widetilde{m} \sim 100$
GeV and $\sin \phi_{A,B} \sim 1$, which violate the experimental bound
by 2 orders of magnitude. This gives rise to the `supersymmetric CP
problem' \cite{buch}.  The contribution to the $\epsilon_K$ parameter
in the neutral $K$ system \cite{masiero} overshoots the experimental
constraint by several orders of magnitude, unless one assumes (i)
heavy and nearly degenerate (first two family) squarks, (ii) near
alignment between quark and squark bases, and (iii) $\sin \phi_{A,B}
\ll 1$.

\v

In fact there is an intricate relationship between flavour violation
and CP violation. Several flavour models, i.e., models with global
horizontal symmetries (abelian \cite{nir-abelian}, non-abelian with
$R$-parity conservation \cite{barbieri}, non-abelian with $R$-parity
violation \cite{gb}), have been constructed, but there is no `the'
flavour model yet!

\section{The Higgs bosons in supersymmetry}
Supersymmetry requires two complex Higgs doublets \cite{hunter}.  Out
of the 8 degrees of freedom they contain, 3 are `eaten up' by $W^\pm$
and $Z$, and the remaining 5 correspond to 5 physical Higgs bosons --
two charged ($H^\pm$) and three neutral. Of the three neutral ones,
one is CP odd ($A$) and two are CP even ($H^0$ and $h^0$). Their tree
level masses are given by
\begin{eqnarray}
m_A^2 &  = & m_{H_u}^2 + m_{H_d}^2 + 2|\mu|^2, \nonumber \\
m_{H^\pm}^2 &  = & m_A^2 + M_W^2, \nonumber \\
m_{h^0}^2 & = & {1\over 2}\left[m_A^2 + M_Z^2 - \sqrt{(m_A^2 + M_Z^2)^2 -
    4m_A^2 M_Z^2 \cos^2 2\beta} \right], \\
m_{H^0}^2 & = & {1\over 2}\left[m_A^2 + M_Z^2 + \sqrt{(m_A^2 + M_Z^2)^2 -
    4m_A^2 M_Z^2 \cos^2 2\beta} \right]. \nonumber
\end{eqnarray}
It immediately follows that $m_{H^\pm} \geq M_W$, $m_{H^0} \geq M_Z$
and $m_{h^0} \leq M_Z$ at tree level. The interesting thing to observe
is the last inequality showing the existence of a light neutral Higgs
boson.  This is not really unexpected as the scalar quartic coupling
in supersymmetry is related to the gauge coupling, unlike in the SM
where it is a free parameter.  But the radiative correction to
$m_{h^0}^2$ grows as the fourth power of the top mass and hence is
quite large \cite{higgs}:
\begin{eqnarray}
\delta m_h^2 =
{\O(\alpha)}{\frac{m_t^4}{M_W^2}}
\ln\left(\frac{m^2_{\widetilde{t}}}{m_t^2}\right).
%\label{}
\end{eqnarray}
The radiative correction pushes the upper limit on $m_{h^0}$ to about
150 GeV. This constitutes a clinching test of supersymmetry. If Higgs
is not found within this limit, supersymmetry in its present form is
ruled out, no matter how heavy one would like the superparticles to
be.

\section{Naturalness criterion}
This is more an aesthetic point of view! When we do not understand the
deeper structure of a theory responsible for the origin of some
parameters used for an effective description of the theory, we do not
generally expect that those parameters can be arbitrarily large such
that a delicate cancellation among them may reproduce a small physical
observable. In other words, a model is less `natural' if it is more
`fine-tuned'.  Let us try to understand the situation in the context
of supersymmetric theories. From the scalar potential minimization, we
obtain
\begin{eqnarray}
\frac{1}{2} M_Z^2 = \frac{m_{H_d}^2 - m_{H_u}^2 \tan^2\beta}
{\tan^2\beta - 1} - \mu^2,
\label{nat}
\end{eqnarray}
where $m_{H_u}^2 = m_{H_d}^2 - \Delta m^2$, where $\Delta m^2$ is the
correction due to renormalization group running from the high scale to
the electroweak scale. This correction crucially depends on the top
quark Yukawa coupling.  EWSB occurs when $m_{H_u}^2$ turns negative by
way of $\Delta m^2$ overtaking $m_{H_d}^2$ such that a cancellation
between the two terms on the RHS of Eq.~(\ref{nat}) exactly reproduces
the experimental $Z$-mass on the LHS of the same equation. Now notice
that this is a cancellation between terms of completely different
origin: the first term on the RHS of Eq.~(\ref{nat}) involves soft
scalar masses which appear in the scalar potential after supersymmetry
breaking, while the second term, i.e. the $\mu$ term, arises as a
result of hidden sector interaction and appears in the superpotential.
How much cancellation between these completely uncorrelated quantities
are we going to tolerate?  Barbieri and Giudice (BG)
\cite{naturalness} offered a criterion by introducing a quantity
\begin{eqnarray}
\Delta_i \equiv
\left|\frac{\partial M_Z^2/M_Z^2}{\partial a_i/a_i}\right|,
\label{delta}
\end{eqnarray}
where $a_i$ are input parameters at high scale. $\Delta$ is a measure
of fine-tuning. An upper limit on $\Delta$ can be translated into an
{\em upper} limit on superparticle masses. BG had shown that with
$\Delta = 10$, i.e., with $1/\Delta = 10\%$ fine-tuning, the upper
limits on superparticle masses turn out to be around 1 TeV in the MSSM
with universal boundary conditions. In GMSB models the naturalness
problem is more serious since the right-handed selectron is
significantly lighter than the Higgs (see, for example, Bhattacharyya
and Romanino in \cite{naturalness}). A detailed analysis by Giusti,
Romanino and Strumia \cite{naturalness} has claimed that only 5\% of
the MSSM parameter space is now experimentally allowed with a modest
naturality requirement.  It should be admitted though that naturalness
upper limits are rather subjective and should not taken as very strict
or rigid limits.

\section{$R$-parity violation in supersymmetry}

Even if $U(1)_R$ symmetry is broken to ensure gaugino mass generation,
a discrete symmetry, called $R$-parity, may survive.  $R$-parity is
defined as $R_p = (-1)^{3B+L+2S}$, where $B$ and $L$ are the baryon
and lepton numbers respectively, and $S$ is the spin of the particle.
All SM particles have $R_p = 1$ and their superpartners have $R_p =
-1$. However, it should be noted that $R_p$ conservation is not
ensured by gauge invariance or any other deep underlying principle.
Hence in a general supersymmetric model $R_p$-violating interactions
\cite{rpar} should be included. Stringent constraints on the strength
of $R_p$-violating interactions arise from the considerations of
proton stability, $n$-$\bar{n}$ oscillation, neutrino-less double beta
decay, charged-current universality, $\Delta m_K$ and $\Delta m_B$,
electroweak precision data from LEP, etc. For a collection of these
limits, see \cite{rp-reviews} (and also the references within
\cite{rp-reviews}).

\v

The explicit $R$-parity violating superpotential is given by
\begin{eqnarray}
W_{\not{R_p}} = \frac{1}{2} \lambda_{ijk} \L_i \L_j \E^c_k +
\lambda'_{ijk} \L_i \Q_j \D^c_k +
\frac{1}{2} \lambda''_{ijk} \U^c_i \D^c_j \D^c_k
+ \mu_i \L_i \H_u.
\label{wrp}
\end{eqnarray}
The first, second and the fourth terms are $L$-violating and the third
is $B$-violating. There are 9 $\lambda$ type, 27 $\lambda'$ type, 9
$\lambda''$ type and 3 $\mu_i$ couplings. We briefly mention some
important consequences of $R_p$ violation:
\begin{itemize}
\item The LSP is no longer stable. So supersymmetry search strategies
have to be redesigned, since the characteristic missing energy
signature is now gone! One has to look for multilepton or multijet
final states. For the impact of $R_p$-violating couplings at various
colliders, see \cite{rp-collider}.

\item Since the LSP is not stable, supersymmetry can no longer provide
  a cold dark matter candidate. One must look for an alternative.

\item It is possible to generate Majorana neutrino masses with the
  $L$-violating couplings both at tree and at one loop level.  The
  trilinear $L$-violating couplings induce one loop neutrino masses
  via fermion-sfermion loops, and the bilinear $L$-violating couplings
  contribute to the tree level mass via neutrino-neutralino mixing.
  The couplings within their experimental limits can explain the
  observed neutrino oscillation data \cite{rp-nu}.

\item Complex $R_p$ violating couplings can induce large CP violation
  in some $B$ decay processes in which the SM predicts very small CP
  violation \cite{rp-b}. These effects can be tested in the ongoing
  and upcoming $B$ factories.

\end{itemize}

\section{Conclusion and Outlook}

Some of the questions that guide our hunt for supersymmetry are:
\begin{itemize}
\item What we expect to be the first signal of supersymmetry? In
which machine? And when?
\item What is the scale of supersymmetry breaking? Is it high or low?
\item Is there any hidden sector? How the information of supersymmetry
  breaking is transmitted from there to the observable world?
\item What is the LSP? Can it account for the cold dark matter?
\item Is $R$-parity violated?
\item Is there a Grand Unification? When will the proton decay?
\item What will be the most convincing solution to the supersymmetric
flavour and CP problems?
\item Why is $\mu$ at the weak scale?
\item Are there extra dimensions?  Do they help supersymmetric model
  building?
\end{itemize}

\v

Only experimental data can uncover the truth.  Supersymmetry has not
been discovered at LEP. But Tevatron is running, LHC is due in 2006,
NLC would hopefully be approved!

\v

We have many models of supersymmetry breaking: SUGRA, GMSB, AMSB,
Gaugino mediation, etc. These models are very predictive and address
the flavour problems differently. They have only a few independent
parameters. But Nature might not have chosen any one of those models
to break supersymmetry!

\v Even though we are not quite sure how supersymmetry is broken, we
know how to parametrize a general supersymmetric theory. The
parameters are then independent. The final answer will come only after
we measure all the 106 supersymmetry parameters (26 masses, 37 angles,
43 phases) in $R$-parity conserving case and a lot more if $R$-parity is
violated. In fact, assuming charginos and neutralinos will be copiously
produced in the NLC, analyses of how to disentangle the CP
violating phases have already started \cite{zerwas}.

\v We make a remark that the large number of parameters in a general
supersymmetric model only reflects our lack of knowledge of the exact
supersymmetry breaking mechanism. As a result, the predictions vary in
a wide range in many cases depending on the choice of parameters. In
this sense, supersymmetry is not simply just one model, it is rather a
class of models.

\section*{Acknowledgments}
I thank the DESY Theory Group, Hamburg, for hospitality, where this
manuscript has been written up. I also thank A. Raychaudhuri for
reading the manuscript.

\end{document}